\documentclass[12pt]{article}

\newcommand{\beq}{\begin{equation}}
\newcommand{\beql}[1]{\begin{equation}\label{eq:#1}}
\newcommand{\eeq}{\end{equation}}
\newcommand{\be}{\begin{equation}}
\newcommand{\ee}{\end{equation}}
\newcommand{\beqn}{\begin{eqnarray}}
\newcommand{\eeqn}{\end{eqnarray}}
\newcommand{\bea}{\begin{eqnarray}}
\newcommand{\eea}{\end{eqnarray}}

\DeclareFixedFont{\xiiss}{OT1}{cmss}{m}{n}{12}
\DeclareFixedFont{\ixss}{OT1}{cmss}{m}{n}{9}
\DeclareFixedFont{\cmrnine}{OT1}{cmr}{m}{n}{9}

\newcommand{\CC}{\hbox{\xiiss C\kern-.4emI}}
\newcommand{\RR}{\hbox{\xiiss R\kern-.45emI}}
\newcommand{\ZZ}{\hbox{\xiiss Z\kern-.4emZ}}
\newcommand{\CCs}{\hbox{\ixss C\kern-.4emI}}
\newcommand{\ZZs}{\hbox{\ixss Z\kern-.4emZ}}
\newcommand{\pa}{\partial}

\newcommand{\pasl}{\pa\kern-.55em /}
\newcommand{\Dsl}{D\kern-.65em /}

\def\href#1#2{#2}

\newcommand{\del}{\partial}
\newcommand{\tI}{\theta^I}
\newcommand{\tJ}{\theta^J}
\newcommand{\ei}[1]{e_i^#1}
\newcommand{\Th}[1]{\bar{\theta}^I\gamma^#1 }
\newcommand {\ha}{ \hat{a}}
\newcommand {\hb}{\hat{b}}

\begin{document}
\begin{titlepage}
\title{ 
        \begin{flushright}
        \begin{small}
        RU-TH-99-10\\
        hep-th/9902046\\
        \end{small}
        \end{flushright}
        \vspace{1.cm}
 On the Quantization of the GS String
on  $AdS_5 \times S^5$}

\author{
Arvind Rajaraman\thanks{e-mail: \tt arvindra@physics.rutgers.edu}
\ and
Moshe Rozali\thanks{e-mail: \tt rozali@physics.rutgers.edu}\\
\\
        \small\it Department of Physics and Astronomy\\
        \small\it Rutgers University\\
        \small\it Piscataway, NJ 08855
}

\maketitle

\begin{abstract}

\end{abstract}
 We discuss the quantization of the Green-Schwarz string action on
 $AdS_5
\times S^5$. We  construct  consistent, globally
 well-defined, gauge
fixing choices for kappa symmetry and worldsheet diffeomorphism
 invariance.
We then proceed to quantize the theory in a perturbation series
in the inverse radius of curvature, in a background field expansion. 
 We discuss vertex operators and correlation functions, and
 demonstrate
agreement with supergravity results in the appropriate limit.
\end{titlepage}


\section{Introduction}

The AdS/CFT correspondence \cite{Maldacena:1997re,witten,gubser}
relates 
string theory
on an $AdS_{d+1} \times S^{p}$ space to a $d$- dimensional
conformal field theory. In particular, type IIB
string theory on $AdS_5\times S^5$ is conjectured to be dual to
N=4 $SU(N)$ Yang-Mills theory in 3+1 dimensions.
In view of this conjecture, it is important to understand
the formulation of string theory on these spaces. 
Such a formulation would provide a nonperturbative
solution of supersymmetric Yang-Mills theory.

 However, formulating string theory on these backgrounds is 
not a straightforward exercise. The
basic reason for this  is that these backgrounds typically
involve a Ramond-Ramond field strength, and such
backgrounds are difficult to describe in the usual RNS
formulation of perturbative
string theory. An exception is the  case of $AdS_3\times S^3$, where
there
exists  a solution with a background NS-NS field strength. This 
 background has been treated in the RNS formulation in
\cite{Giveon:1998ns,Kutasov:1998zh}.

Another approach to string quantization in  $AdS_{d+1} \times S^{p}$
backgrounds is the Green-Schwarz (GS) formalism \cite{gsw}. Indeed, 
actions which represent the classical
GS string in these backgrounds have been found. These actions can be 
written for both the
$AdS_5\times S^5$ background \cite{mt,Kallosh:1998zx,kr,kt,Pesando:1998fv}
 and the $AdS_3\times S^3$
background \cite{ads3,pesando,rey}.
The
reason that the GS approach may be appropriate is that spacetime
supersymmetry is manifest, and so there is no 
fundamental difference between NS-NS and R-R backgrounds. 

These actions are, however, only classical, and need
to be quantized. In particular, there are various
gauge symmetries, of reparametrization and kappa-symmetry,
 that need to be  fixed.
In the case of the flat space GS action,
this is most effectively done in light-cone gauge \cite{gsw}.
However, the backgrounds under consideration have
no globally defined light-like direction, and therefore this gauge is
invalid.
Much of our discussion therefore, revolves about
choosing a gauge-fixing. This is described in section 3.

After the gauge fixing, the action does not simplify
very much; it is still a complicated sigma-model
with several interaction terms. It can nevertheless
be treated in sigma-model perturbation theory,
since (as we show) there is a well defined expansion
in powers of the curvature of the background. This
is related by the AdS/CFT correspondence to a 
perturbation series in ${1\over g^2N}$ in the
dual Yang-Mills theory.

We  present leading order expressions for
vertex operators in the theory. These expressions
provide a starting point for constructing the
vertex operators in powers of the curvature.
Similarly, correlation functions can be computed
in a perturbation expansion. We show
that the leading order correlation functions
 reduce to previously computed results. In particular 
we recover the supergravity procedure of calculating the 
CFT correlation functions \cite{witten,gubser}.

One should note that we do not prove that the sigma-model we consider
defines a spacetime
 conformal field theory.
There have been arguments that this is the case
\cite{Kallosh:1998qs,mt}, but
a direct proof is still lacking. The consistency of our 
results is, however, an encouraging sign.

\section{The Green-Schwarz String action}

We review here the Green-Schwarz string action constructed in
\cite{mt}.
 We concentrate on the case of type IIB string theory on $AdS_5 \times
S^5$, but the discussion can also be applied to similar string backgrounds,
such as the Green-Schwarz string on
  $AdS_3 \times S^3$, discussed in \cite{ads3,pesando,rey}.

We shall use the metric of $AdS$ space in the form
\be
ds^2=-(1+e^2r^2)dt^2+{dr^2\over 1+e^2r^2}+r^2d\Omega_3^2
\,
\ee
 where $e^2$ is the cosmological constant in string units, related to
gauge theory parameters as $\frac{1} {e^4} =  N {g^2_{YM}}$.
 
The action is constructed as a $\sigma$-model action with the target
space
being the coset $\frac{SU(2,2|4)}{SO(4,1)\times SO(5)}$. The action is
written in terms of the supervielbeins $L^{a}, L^I$, and is given as :
\begin{eqnarray}
S =-\frac{1}{2}\int_{\partial {M_3}}  d^2\sigma\  \sqrt{g} \, g^{ij}
 L_i^{\hat a} L_j^{\hat a} +  {\rm i}\int_{M_3}
 s^{IJ}   L^{\hat a} \wedge  \bar{L}^I\gamma^{\hat a}\wedge  L^J
\ ,
\label{action}
\end{eqnarray}
 where $\partial {M_3}$ is the string worldsheet.

The supervielbeins, viewed as 1-forms over the coset manifold, are
restricted to satisfy the Maurer-Cartan equations, given explicitly
in \cite{mt}. The solution of these equations was found in \cite{mt}.
The expressions for the supervielbeins
 $L^{a}, L^I$ are given  as :
\begin{equation}
L^{ I}  =\left[ \left({\sinh {\cal M} \over {\cal M}}\right) D\theta \right]^{I}
\label{I}
\end{equation}
and
\begin{eqnarray}
L^{\hat a }&=&e^{\hat a }_{\mu} (x) dx^{\mu}   -4 i  \bar \theta^I \gamma^{\hat
a}
\left({
\sinh^2  {\cal M}/2 \over {\cal M}^2} D\theta \right)^I\label{La}
\end{eqnarray}
where ${\cal M}^2$ is given by 
\begin{eqnarray}
({\cal M}^2)^{ I} _{ L}&=& e^2 [ \epsilon^{IJ}
(-\gamma^{ a}  \theta^{J}(x) \bar \theta^L(x) \gamma^{ a} + \gamma^{ a'}
\theta^{J}(x) \bar \theta^L(x) \gamma^{ a'} )\nonumber\\
&+& {1\over 2}
\epsilon^{KL} (\gamma^{ab} \theta^I (x)\bar \theta^K (x)\gamma^{ab}
+\gamma^{a'b'} \theta^I(x) \bar \theta^K(x) \gamma^{a'b'})] \label{msquare}
\end{eqnarray}

Here,
\begin{equation}
(D\theta)^I  =\left ( d  +{1\over 4}(  \omega^{ab} \gamma_{ab} +
\omega^{a'b'}
\gamma_{a'b'}) \right )\theta^I  -{e\over 2} i\epsilon^{IJ} (e^a
\gamma_a +
ie^{a'} \gamma_{a'}) \theta ^J
\end{equation}

Our notation is that of \cite{mt}; 
$a=0,1,2,3,4$ are tangent space indices  for $AdS_5$ and $a'=
5,6,7,8,9$ the corresponding indices 
 for $S^5$. The index ${\hat a }$ runs over $0,1..9$, and  $e^{\hat a }$,
$\omega^{{\hat a}{\hat b}}$ are the bosonic vielbein and the spin
connection
 respectively.

 We work in a Euclidean worldsheet, and fix the worldsheet metric to
be
$g_{ij}= \delta_{ij}$. In this gauge,
the equations of motion for the sigma model are \cite{mt}
\begin{eqnarray}\label{eqqs}
&&
\delta^{ij}(\partial_i L_j^a+L_i^{ab}L_j^b)
+{\rm i}\epsilon^{ij}s^{IJ}\bar{L}_i^I\gamma^a L_j^J
=0\,,
\\
&&
\delta^{ij}(\partial_i L_j^{a'}+L_i^{a'b'}L_j^{b'})
-\epsilon^{ij}s^{IJ}\bar{L}_i^I\gamma^{a'} L_j^J
=0\,, \\
&&
(\gamma^a L_i^a+{\rm i}\gamma^{a'} L_i^{a'}) (\delta^{ij} \delta^{IJ} - \epsilon^{ij}
s^{IJ})       L_j^J=0\,,
\end{eqnarray}

As a result of the gauge fixing,
these relations should be supplemented  by   the standard Virasoro
constraints
\be L^a_i L^a_j+L^{a'}_i L^{a'}_j = {1\over 2} \delta_{ij} 
 (L^a_l L^a_l+L^{a'}_l L^{a'}_l) \ ,
\ee

 In this paper we are interested in quantizing the action in leading
order in the cosmological constant $e^2$. The supervielbeins reduce in
the leading
order to:
\begin{eqnarray}
L^{\ha}&=& e^{\ha} + i \theta^I \Gamma^{\ha} D\theta^I \nonumber \\
       L^I &=& D\theta^I
\end{eqnarray}

 These expressions are the  covariant extensions of the corresponding
expressions of  the  flat
space Green-Schwarz string. In particular, the interaction with
the background curvature, and the interaction with the RR background, 
are both included in the leading order action. 

This  leading order action, to be discussed in the rest of the paper, is
given in \cite{mt}. It is the covariantization of the usual flat space
GS
action:
\begin{equation}
\label{action1}
{\cal L}_1= -{1\over 2}({e_i}^{\ha} -i\Th{{\ha}} D_i{\tI})^2
-i\epsilon_{ij} {e_i}^{\ha}
s^{IJ}\Th{{\ha}} D_j\tJ +\epsilon^{ij}(\bar{\theta}^1\gamma^{\ha} D_i\theta^1)
(\bar{\theta}^2\gamma^{\ha} D_i\theta^2).
\end{equation}

We wish to quantize this action in a background field expansion. The
background we choose is a worldsheet at a constant spatial
position. The 
gauge we wish to work in is the
covariantization
of the flat space $X_0 =\tau$ gauge. In order to perform this
quantization
we need to define our gauge choices globally, before going to any 
limit of interest. We turn, then, to discuss the gauge fixing of the
action.

\section{Gauge Fixing}

In order to quantize the action we first need to 
gauge-fix the worldsheet gauge symmetries : reparametrization invariance 
and kappa-symmetry.  

We first change the fermionic basis to the ``Killing spinor
basis''\footnote{ This is a somewhat misleading term, as  $\Theta^I$
are  generally not Killing spinors. However, we stick to the established
notations.}
\cite{kr}. These, by definition, satisfy
\bea
(\Theta^{\tilde{\alpha}})^I &=& {E_{\alpha}}^{\tilde{\alpha}} 
(\theta^{\alpha})^I \\
(d\Theta^{\tilde{\alpha}})^I &=&  {E_{\alpha}}^{\tilde{\alpha}} 
(D\theta^{\alpha})^I
\eea
These equations permit solutions because of the highly
symmetric nature of the space $AdS_5\times S^5$, which
has 32 Killing spinors. We note also that the covariant derivative
used in this definition includes an interaction with the  RR
background, in 
addition to
the usual coupling to the spin connection.

In this basis, the equations of motion for the spinors simplify
\be
\label{eom}
(\delta_{ij}\delta^{IJ}- \epsilon_{ij} S^{IJ})\partial_j \Theta^I =0
\ee

We can now fix the kappa symmetry. Denote
$\Gamma^+ = \Gamma ^0 + \Gamma^r$, and choose :
\begin{equation}
\Gamma^+ \Theta^I =0
\end{equation} 
This gauge is globally well-defined, unlike the gauge choices
of \cite{kr,kt}. The reason for imposing this gauge on the
Killing spinor basis rather than the original basis is
that the original fermions  pick up
a phase factor if the string is transported around a
path with nontrivial holonomy. The spinors $\Theta^I$, on
the other hand, do not pick up such a factor. This can
be seen from the fact that their kinetic term does not
contain  a coupling to the spin connection or to the RR background. Hence this
gauge choice is globally well defined.

We now turn to fix the 
 worldsheet diffeomerphism invariance.  We  already chose the
conformal gauge:
$g_{ij}=\delta_{ij}$.  
 The conformal gauge leaves an unbroken subgroup of the worldsheet
gauge symmetries, the conformal transformations.  A consistent way to
fix those residual gauge symmetries  is described in \cite{gsw}. One
may  choose any free field on the worldsheet and set it to equal one of
the
worldsheet coordinates, say $\tau$. This consistently
 forces (almost) all the Fourier modes of the gauge parameter
to vanish, leaving  $\sigma$ rigid translations, generated by
$L_0 - {\bar L_0}$, as the only leftover
 gauge freedom.  This gauge
choice is utilized in the
flat space string action to fix the lightcone gauge, as discussed in
\cite{gsw}.

 Similarly, one may utilize an exactly conserved current on the
worldsheet to fix the conformal transformation. Being exactly conserved
 determines the conformal transformation law of the current.
 Therefore, fixing $J_i$ to be a constant forces almost all of the Fourier
modes of the gauge parameter to vanish. The leftover gauge invariance
in
this case consists of rigid translations on the worldsheet, generated
by
both $L_0$ and ${\bar L_0}$.

We wish to choose a gauge which reduces to the temporal
gauge $X_0 = \Delta \tau$
asymptotically
in spacetime (where $\Delta$ is a constant). As a timelike 
coordinate can be defined globally, unlike
a null coordinate, this gauge choice can be made globally
well-defined.
Moreover, the momentum conjugate to the global time defines
the dimension operator in the spacetime CFT. Therefore the 
constant $\Delta$ has a natural interpretation in that CFT.

  The reader may be aware of the notorious problems 
of string quantization in the temporal gauge, in the flat space case.
The interpretation of  the quantization procedure in the case of the
$AdS$
space is, however, radically different. For example, by 
constructing vertex operators of the first quantized string one 
constructs operators in the spacetime CFT. Those operators are not
required to have a particle interpretation, and indeed lie generally in
non-unitary representations of the spacetime symmetry group.
 Furthermore, the dual 
gauge theory is believed to be consistent when treated in the  
global time, having 
a Hamiltonian and unitary time evolution\footnote{ We thank Tom Banks for
  discussions on this point.}. 

 To work in the temporal gauge it is natural to utilize the time
 translation current on the worldsheet:

\beq
 {J_i}^0=e^0_0 \left(\ei{0}+i\Th{0}D_i\tI \right) + \mathcal{O}(e^2)
\eeq
 
In order to simplify the action it is convenient to add to this current
 another  conserved current which has no action on the zero modes of the
 fields. It therefore corresponds to a uniform shift of the dimension
of each multiplet separately. The shift can be shown to vanish
for the supergravity vertex operators constructed below\footnote{ The
 shift is given by a certain 
  number operator for fermionic oscillators.}.

The conserved current we use to fix the gauge is then:
\beq
J_i = {J_i}^0 - i(\Theta^I)^\dagger\del_i \Theta^I
\eeq

 The extra current is exactly conserved as a consequence of the 
equation of motion (\ref{eom}). With this gauge fixing taken the 
fermions have a non-degenerate kinetic term. 

Thus, in order to specify the temporal gauge we would like to 
choose:

\beq
J_i = \Delta \delta_{i \tau}
\eeq

However, this is not possible, as it is inconsistent with the
equations of motion\footnote{We thank E. Witten for pointing this  out to
us.}.
 Indeed, the current $J_i$ satisfies:
\bea
\partial_0 J_0 + \partial_1 J_1 &=& 0 \\
\partial_0 J_1 - \partial_1 J_0 &=& \mathcal{O}(e^2)
\eea

This implies:
\be
\label{cor}
\left({\partial_0}^2 -{\partial_1}^2 \right) J_i = \mathcal{O}(e^2) 
\ee

which is only consistent with our gauge choice to leading order in
$e$. At higher orders in $e$ we must correct the gauge choice
perturbatively.
This can be done by adding the inhomogeneous solution of (\ref{cor})
to the gauge choice above.

The constant $\Delta$ is to be interpreted as the total dimension, 
similar to the total light cone momentum $P^+$ in lightcone
quantization. This can be demonstrated for supergravity states  as follows.

Define the worldsheet current corresponding to $X^{\mu}$ translation 
as $J^{\mu}_i$. Assuming the spacetime superconformal algebra is
an exact symmetry of the quantum theory, one can work in states that
are representations of the superconformal algebra. Therefore
\beq
(\int d\tau J^{\mu}_{\tau})|\mbox{ state}> = \cal{R}^{\mu} |\mbox{ state}>
\eeq
where $\cal{R}^{\mu}$ is a generator in 
some matrix representation of the spacetime 
superconformal algebra.

We take now the mini-superspace approximation, setting all $\sigma$
derivatives to zero. This is the particle limit of the string theory,
and the states in this approximation are those of the first quantized
supergravity. The worldsheet current satisfies:

\beq
\partial_{\tau}J^{\mu}_{\tau} = 0
\eeq

 In addition one has the Virasoro constraint:
\beq
T_{00} = G_{\mu\nu}J^{\mu}_{\tau}J^{\nu}_{\tau} =0
\eeq

This constraint sets the second Casimir of the superconformal algebra,
in the  $\cal{R}$ representation, to zero. This is exactly the 
equation satisfied by the supergravity modes, which determines the
dimension of any mode in terms of its other quantum numbers.

In the gauge chosen this
equation relates the constant $\Delta$ to the quantum numbers of the state
in exactly the same way (up to a trivial shift). This result
is independent of the approximation we take next, the background field
expansion.

With the same assumption, namely that the quantum theory preserves the 
spacetime superconformal algebra, one can organize  the string states
into supersymmetry multiplets. We note that as a consequence of the
superconformal algebra, members of any given multiplet carry different
dimensions. Specifically, this follows from the commutation relations:

\be
\left[ Q,D \right] = {Q \over 2} 
\ee
where $Q$ are the supersymmetry generators, and $D$ is the dimension. 

 Formally, therefore, different states in the same supersymmetry multiplet
belong to separate Hilbert spaces. In the leading order approximation
around flat
space, however, the splitting is zero, and multiplets are degenerate. 
A proof of closure of the spacetime symmetry algebra to the
first subleading order would  therefore  establish the required
 splitting.

\section{Background Field Expansion}

 Having defined our  gauge choices globally, we are ready to 
use the resulting action. We proceed by expanding the action
in a background field expansion 
around the following worldsheet configuration:

Define $ X^{\mu}=X^{\mu}_0+\xi^{\mu}$, where $X^{\mu}_0$ are constants, and
${\mu}=0,...,9$ is a curved coordinate index. In order to normalize
the fluctuating fields canonically define:
\be
\label{rescaling}
\xi^{\ha} = e^{\ha}_{\mu}(X^{\mu}_0)\xi^{\mu}
\ee
Where $\ha=0,...,9$ is a tangent space index.

 We expand the  action (\ref{action1}) in
powers of $\xi^{\ha}$, the bosonic fluctuations.
Since $r$ determines the magnitude of spacetime derivatives,
this is an expansion in $\xi/r$. Thus, this approximation is useful
in  studying small fluctuations of a worldsheet located asymptotically in 
spacetime, near the boundary of the $AdS$ space. In particular it is
suited for the study of non-normalizable modes, which correspond to
the operator algebra of the spacetime CFT.

 The action and the gauge choice simplify significantly in the leading
order in the background field expansion, becoming essentially 
the flat space GS action in the $X^0 = \tau$ gauge. However,  the treatment of
the zero modes, bosonic and fermionic, is sensitive to the 
curved nature of the space.

 The action (\ref{action1}) is constructed from the 
following ingredients, written
to first subleading order in $e$, the cosmological constant. We use
the indices $ m,n,... = 1,...,8$ to denote the ``transverse'' fluctuations. 
The kappa symmetry gauge choice is imposed, but we delay imposing the 
diffeomeorphism gauge choice till later.
\bea
\label{expansion}
  e_i^{\ha} &=& \del_i \xi^{\ha}  + B_{bc}^a(X_0) \xi^b \partial_i \xi^c   \\
D_i\tI &=& \partial_i\tI+
{1\over 4}(  \omega_i^{\ha \hb}(x_0) \gamma_{\ha \hb} \tI
 -{e\over 2} i\epsilon^{IJ} e^{\ha}_i(x_0)
\gamma_{\ha} )
 \theta ^J = \nonumber\\
&=& \partial_i\tI - \frac{e}{2} \partial_i \xi^m \gamma^{mr}\tI -{e\over
2} 
i\epsilon^{IJ} (\partial_i \xi^{\ha} \gamma^{\ha})
 \theta ^J 
\eea
where
\be
 B_{bc}^a = e^\nu_b e^\mu_c \partial_\nu e_\mu^a
\ee

The action to leading order in $e$, before gauge fixing, is 
the flat space GS action
\be
{\cal L}_1=-{1\over 2}(\partial_i \xi^{\ha} -i\Th{{\ha}} \partial_i\tI)^2
-i\epsilon_{ij}\partial_i \xi^{\ha}
s^{IJ}\Th{{\ha}} \partial_j\tJ +\epsilon^{ij}(\bar{\theta}^1\gamma^{\ha}
\partial_i\theta^1)
(\bar{\theta}^2\gamma^{\ha} \partial_i\theta^2).
\ee

The gauge fixing is
\be
J_i=e^0_0(x_0)(\ei{0}+i\Th{0}D_i\tI)-i({\Theta}^I)^\dagger\del_i{\Theta}^I
\ee
 
 In the limit taken, this expression reduces to:

\bea
J_i=e^0_0(x_0)\partial_i \xi^0
\eea
 Though the gauge choice has to be imposed exactly, approximating
$J_i$ by its leading order 
changes the action only in subleading terms in the background field
expansion. To find the complete  action to first subleading order one
needs to include the next order action written in \cite{mt}, use the
expansion (\ref{expansion}), and expand the gauge choice as well.
  
Thus we impose in the leading order:

\beq
e^0_0(x_0)\partial_i \xi^0= \delta_{i\tau}
\eeq

In this limit the action for the fluctuations simplifies. The 
transverse fluctuations and the fermions obey free field equations,
and can be expanded in modes. The zero mode structure is still sensitive
to the
global structure of the background. We demonstrate this by
constructing
some simple vertex operators in the next section.

\section{Vertex Operators}

In order to find the vertex operators one needs to solve the Virasoro
constraints. Apart from the zero-mode contribution, this is identical
to solving the constraints in flat space, in the $X^0 =\tau$ gauge. 

We use the following notation: we work in worldsheet lightcone
coordinates $\sigma_{\pm} = \sigma \pm \tau$. We concentrate on the
left-moving 
part, and suppress the corresponding  indices.  In spacetime we
use
the indices $a,b,..$ to denote all coordinate labels except for $0$,
the gauge fixed coordinate.

The fermion
$\Theta^1$ is left-moving by its equation of motion, and we denote it
by $S$.

The  left moving Virasoro constraint is found to be:

\beq
T = (\partial \xi^m)^2 - (\frac{\Delta}{e_0^0(X_0)} - i \bar{S} \gamma^0
\partial S)^2 + (\partial \xi^r + i \bar{S} \gamma^0
\partial S)^2
\eeq

 The bosonic fluctuations have the mode expansion:

\beq
\xi^a = e^a_{\mu}(X_0)(\xi^{\mu})_0 + \sum_{n \neq 0} \frac{1}{n}\alpha_n^a \exp(in
\sigma_+) + \mbox{right movers}
\eeq

  The normalizations chosen in this mode expansion need to be
  clarified. The fundamental variables to be quantized are the unrescaled
  fluctuations
$\xi^{\mu}$ defined above. For oscillator modes the rescaling
in equation (\ref{rescaling}) has the mere effect of 
changing their normalization.
 However, the zero mode of $\xi^r$ does not commute with
  $e^a_{\mu}(X_0)$,
hence we choose to keep the original zero modes $(\xi^{\mu})_0$ , and
  their
conjugate momenta $P^{\nu}$, as our canonical variables.

 One
imposes the following commutaiton relations:
\bea
\left[ (\xi^{\mu})_0, P^{\nu} \right]&=& - G^{\mu\nu}(X_0)\nonumber\\
\left[ \alpha_n^a, \alpha_m^b \right] &=& i m \delta_{m+n,0} 
\eea

 For the fermions one has the mode expansion:
\beq 
S = S_0 +  \sum_{n \neq 0} S_n \exp(in
\sigma_+) 
\eeq
and the anti-commutation relations are:

\beq
 \left\{ (S_m)^\dagger, S_n \right\} = i \gamma^0 \delta_{m+n,0} 
\eeq

 Define $L_n$ to be the Fourier modes of $T$.
The gauge fixing chosen here fixes almost all the conformal transformations,
leaving unfixed the rigid translations 
generated by $L_0$ (and $\bar{L_0}$). Thus one
needs
to solve the constraints for all $L_n$, $n \neq 0$. These can be used
to
express all the modes $\{ \alpha_n^r \}$ in terms of the transverse
fluctuations
$\xi^m$. 

 In the absence of fermionic  oscillators these equations simplify,
 and are given as:

\beq
\label{virasoro}
L_n =  \sum_{k} \alpha_{n-k}^a \, \alpha_k^b =0
\eeq

These are quadratic equations for the modes $\{ \alpha_n^r \}$, unlike the 
linear equation in the lightcone gauge quantization of the flat
space string. The equation can be formally solved as:

\beq
\label{sol}
\alpha_k^r = \int d\sigma_- \exp(-ik \sigma_-) \sqrt{\Delta^2
G^{00}-(\partial
\xi^m)^2}
\eeq

 The vertex operators represent emission vertices in flat space, and
 their
interpretation in our case is demonstrated by computing their
 correlation function in the next section.  As the vertex operators
carry non-zero dimensions, they cannot be interpreted as acting on the
 Hilbert
space of states with a fixed total dimension $\Delta$, denoted as 
$\mathcal{H}_{\Delta}$. They should be formulated as operators acting 
in a larger Hilbert space, where the dimension $\Delta$
is allowed to change.

  To formally
 express
this fact we introduce the operators  $O_\delta =exp(\delta
 X^0_0)$. Here 
$X^0_0$ is the zero mode of the time 
coordinate $X^0$, unfixed by our gauge choice.  These operators
act naturally in an enlarged Hilbert space $\cal{H}$ defined as the direct 
sum of all the Hilbert spaces $\mathcal{H}_{\Delta}$ of fixed $\Delta$.

 It is convenient also to introduce an operator $\hat{\Delta}$, which
is proportional to the unit operator  in any summand Hilbert 
space $\mathcal{H}_\Delta$ , and takes 
the value $\Delta$ there. Then one has
 the following commutation relation:

\be
\left[\hat{\Delta}, O_\delta \right] = \delta O_\delta
\ee

The vertex operators are functionals of the transverse fluctuations and
their
conjugate momenta, as well as  $r_0$, the radial zero mode, and its 
conjugate momentum. They include an appropriate operator $O_\delta$
 defined above. As a result of the leftover gauge freedom, the 
vertex operators  are restricted to 
satisfy the $L_0$
constraint:
\bea
\left[L_0, V \right] &=& V\nonumber\\
L_0 &=& G^{\mu\nu}(X_0) P_{\mu} P_{\nu} + 
 G^{00}(X_0) {\hat{\Delta}}^2 +\sum_{k} \alpha_{-k}^a \alpha_k^a 
\eea
where in this equation the oscillators $\alpha_n^r$ are given in (\ref{sol}),
 as solutions to equation (\ref{virasoro}). 

As usual one needs to choose operator ordering when writing the
expression for $L_0$, since the momenta $P_r$ do not
commute with the zero mode $r_0$. 
 Note that in this order there is no normal ordering constant in this
expression, as we have exact cancellation between free bosonic and fermionic 
fluctuating modes.

The simplest vertex operators are the supergravity modes. Their
oscillator part is identical to the flat space case, and their 
zero mode wavefunction is determined by the $L_0$ constraint.
 This simply sets the spacetime covariant Laplacian, which is the
second Casimir of the conformal algebra, equal to the ten dimensional
mass of the string state, zero in this case.

For example consider the  vertex operator of a graviton with transverse
 polarization. We also set the transverse momenta to zero, as all
 other cases can be simply obtained from this one.
 The corresponding
 vertex operator
is then:

\beq
\label{graviton}
V^{mn}= \partial X^m \bar{\partial}X^n  \,
O_\delta \, 
\Psi_{mn}(r_0)
\eeq

The $L_0$ constraint translates then to the following equation
on the wave function
$\Psi_{mn}(r_0)$:

\be
G^{00}{\delta}^2 \Psi_{mn}+ \partial_r 
\left( G^{rr} \partial_r
\Psi_{mn} \right) =0
\ee
which is identical to the equation satisfied by the corresponding
supergravity
mode. In order to make correlation functions non-vanishing, one is 
forced to take the non-normalizable solutions 
of this equation, as is discussed in the next section.

\section{Correlation Functions}

 The $AdS$/$CFT$ correspondence includes a prescription for calculating
 correlation functions of CFT operators in the supergravity limit 
of the $AdS$ theory\cite{witten, gubser}. Define the generating
 functional of the $CFT$ correlation functions as $W[\Phi_0]$.  The
 currents
$\Phi_0$ couple to $CFT$ operators, and differentiation of $W[\Phi_0]$
with respect to those currents
yields the correlation functions of those operators.

 The $AdS$/$CFT$ correspondence identifies this generating functional
with the spacetime partition function with  prescribed boundary
conditions.
In the supergravity limit this is simply:
\bea
\label{gen0}
W[\Phi_0]= exp(S[\Phi,g_{\mu\nu}])
\eea

The currents $\Phi_0$ enter the supergravity calculation as boundary
conditions for the supergravity fields, or 
equivalently as deformations of these fields 
by non-fluctuating, non-normalizable modes\cite{lawrence}. 

 We wish to recover this prescription in the present context. In
 string theory one computes 
 correlation functions of vertex operators as:

\be
\left<\prod{ V_i}\right>=\int DX D\theta \ exp(-S)\prod{ V_i}
\ee

These correlation functions are obtained by differentiation of the
following generating functional:

\be
\label{gen}
W[J]=\int DX D\theta \ exp(-S+ \sum_i J_i V_i)
\ee

The expression is readily recognized as the partition function of the string
with the string  action deformed by the addition of 
$
\sum_i J_i V_i$. For the gravity modes, this can be
interpreted
as deformation of the spacetime metric (and its supersymmetric
partners).
 Take for example the graviton vertex operator constructed in the last
section, equation (\ref{graviton}).
Adding it to the string action results in deforming  the spacetime
metric by $O_\delta \Psi_{mn}(r_0)$. 

Therefore we are motivated to 
identify the string vertex operators with the non-normalizable modes
of spacetime fields, and the string generating functional (\ref{gen})
 with the CFT generating functional as in (\ref{gen0}).

Thus, a natural extension of the supergravity prescription
in (\ref{gen0}) is to identify the generating
functional of the CFT correlation functions with the string 
partition function, with
the string action deformed by the corresponding (1,1) operators.

 This leads to
the following identification \cite{ooguri}:

\be
 \left< \prod {O_i} \right> = \left< \prod V_i \right> 
\ee

 Here $V_i$ is the string vertex operator representing the
 non-normalizable
mode which couples to the CFT operator $O_i$.

For the supergravity modes, in the mini-superspace approximation, this
can be shown to reduce to a supergravity calculation. 
In this limit 
the correlation of vertex operators is reduced to
\bea
\label{corr}
\left<\prod{ V_i}\right>=\int DX DS_0 \ exp(-\int d\tau L_0)\prod{ V_i}
\eea
where the string action is (in the minisuperspace approximation)
\bea
L_0=g_{\mu\nu}\partial_\tau X^\mu\partial_\tau X^\nu 
\eea

 We demonstrate the calculation of correlation functions by calculating
the  two point function of
 the graviton vertex operator above, equation (\ref{graviton}), which has
no angular
momentum on either the $S^3$ or the $S^5$.  The vertex operator
is then:
\bea
V_{mn}=\dot{X}^m\dot{X}^n \Psi_{mn}(X)
\eea
where $\Psi_{mn}(X,x)$ is the bulk-to-boundary propagator \cite{witten},
corresponding to an operator insertion at point $x$ on the boundary.

The integral (\ref{corr}) is then
\bea
\int DX DS_0 \ exp\left(-\int d\tau L_0\right)\ \dot{X}^m\dot{X}^n
\Psi_{mn}
(X,x)\ \dot{X}^m\dot{X}^n \Psi_{mn}(X,y)
\eea
we can exponentiate the vertex operators to
\bea
I[J]=\int DX DS_0 \ exp\left(-\int d\tau L_0+J(x)\dot{X}^m\dot{X}^n
 \Psi_{mn}(X,x)\right)
\eea
We can then generate the correlation functions by differentiating
with respect to  $J(x)$.

If $J(x)$ was zero, then the integral equals (see for example \cite{strassler})
\bea
W[0]&=&\int DX DS_0 \ exp\left(-\int d\tau L_0\right) \nonumber\\
 &=& \int DXDp \ exp(p\dot{x})\ exp(-g^{\mu\nu}p_\mu p_\nu)
\eea
which by standard path integral manipulations is
\bea
W[0]=\sum_n exp(-p_n^2T)
\eea
where $p_n$ are the eigenvalues of
\bea
\del_i[g^{ij}\del_j\phi]=p_n^2\phi
\eea
We can therefore replace the sum over modes
by an integral over field configurations
\bea
W[0]=\int D\phi \ exp\ (-g^{\mu\nu}\del_\mu \phi\del_\nu \phi)
\eea

We can redo this calculation in the presence of vertex
operators. It is straightforward to show that \cite{strassler}
\bea
\int DX DS_0 \ exp\left(-\int d\tau L_0+J(x)\dot{X}^m\dot{X}^n \Psi_{mn}(X,x)\right)\nonumber
\\
=\int D\phi \ exp(-g^{\mu\nu}\del_\mu \phi\del_\nu \phi-
J g^{\mu\nu}\del_\mu \phi\del_\nu \psi)
\eea
To calculate a two point function, we differentiate with respect to
$J$ twice.

At this point, it is clear that we have reduced the
computation to the corresponding supergravity calculation, and
thus shown the equivalence of the formulae (\ref{gen0}) and
(\ref{gen}), in the appropriate limit.  

 In order to compare 
to the supergravity result we still have to discuss the 
normalization of the vertex operators.

 In order to normalize correctly the vertex operator (\ref{graviton}),
 one
 needs to
consider their
CFT interpretation. Usually a vertex operator insertion 
carries a factor of $g_{str}$, the string coupling.  This gives, for
 example, a  $g_{str}$ independent result for the two point functions
 calculated on the sphere , as sphere amplitudes are proportional to
$\frac{1}{{g_{str}}^2}$. The three point function on the sphere is
 then proportional to $g_{str}$.

In the present context, correlation functions of the graviton 
are interpreted as correlation functions of the energy-momentum tensor
of the spacetime CFT. As such, they obey conformal Ward identities
which
are $g_{str}$ independent. For example, one such identity
relates the two-point and 3-point functions as follows:
\be
\partial_\mu \left< T^\mu_{\ \nu(x)} T(y) T(z) \right> =
-  \delta(x-y)\partial_{y^\nu}\left<T(y)T(z) \right>
 - \delta(x-z)\partial_{z^\nu}\left<T(y)T(z) \right>
\ee

Thus, the graviton vertex operator cannot carry any factor of
$g_{str}$.
 Consequently, all correlation functions on the sphere are
proportional to $\frac{1}{{g_{str}}^2}$. Additional factors of
$e^2$ may arise from the worldsheet calculation, hence the general
correlation function on the sphere is proportional to
$N^b {g_{str}}^{b-2}$. The constant $b$ is determined
by dimensional analysis, $4b$ being the total conformal dimension
of the correlation function.

The two-point function above is then found to be:

\be
\left<V_{mn}V_{mn} \right> = N^2 \frac{1}{{|x-y|}^4}
\ee

This result agrees with the calculation of the central charge of the
gauge
theory \cite{gubser}.

\section{Acknowledgments}  

We thank O. Aharony, M. Berkooz, N.
Berkovits, M.
Douglas, W. Fischler, L. Susskind, E. Witten and especially T. Banks
for useful conversations.

\bibliography{gs5}
\bibliographystyle{utphys}

\end{document}